\documentclass[twocolumn,tighten]{aastex61}

\usepackage{natbib}
\shorttitle{\hcn\ and \hcop\ in IC 10}
\shortauthors{Kepley et al.} 

\newcommand{\ha}{\ensuremath{{\rm H}\alpha}}

\newcommand{\kms}{\ensuremath{\rm{km \, s}^{-1}}}

\newcommand{\Lsun}{\ensuremath{L_\odot\ }}

\newcommand{\co}[3]{\ensuremath{\rm{^{#1} CO ({#2}-{#3})}}}

\newcommand{\Kkmspc}{\ensuremath{\rm{K \, km \, s^{-1} pc^2}}}

\newcommand{\hcn} {\ensuremath{{\rm HCN}}}
\newcommand{\hcop}{\ensuremath{{\rm HCO^+}}}
\newcommand{\LIR}{\ensuremath{{L_{\rm IR}}}}
\newcommand{\LIRCO}{\ensuremath{{L_{\rm IR}/L_{\rm CO}}}}
\newcommand{\LIRHCN}{\ensuremath{{L_{\rm IR}/L_{\rm HCN}}}}
\newcommand{\LIRHCOP}{\ensuremath{{L_{\rm IR}/L_{\rm HCO+}}}}

\newcommand{\LHCN}{\ensuremath{{L_{\rm HCN}}}}
\newcommand{\LHCOP}{\ensuremath{{L_{\rm HCO^+}}}}

\newcommand{\LHCNCO}{\ensuremath{L_{\rm HCN}/L_{\rm CO}}}

\newcommand{\LHCOPCO}{\ensuremath{L_{\rm HCO^+}/L_{\rm CO}}}

\newcommand{\IHCNHCOP}{\ensuremath{I_{\rm HCN}/I_{\rm HCO^+}}}
\newcommand{\coh}{CO-to-H$_2$}

\newcommand{\coone}{\co{12}{1}{0}}
\newcommand{\fdense}{\ensuremath{f_{dense}}}
\newcommand{\SIR}{\ensuremath{S_{\rm IR}}}

\begin{document}


\title{Dense Molecular Gas in the Nearby Low Metallicity Dwarf Starburst Galaxy IC 10}

\correspondingauthor{A. Kepley}
\email{akepley@nrao.edu}

\author{Amanda A. Kepley}
\affiliation{National Radio Astronomy Observatory, 520 Edgemont Road, Charlottesville, VA 22903-2475, USA}

\author{Lauren Bittle}
\affiliation{Department of Astronomy, University of Virginia, 520 McCormick Road, Charlottesville, VA 22904-4325, USA}

\author{Adam K. Leroy}
\affiliation{Department of Astronomy, The Ohio State University, 140 West 18th Avenue, Columbus, OH 43210, USA}

\author{Mar\'ia J. Jim\'enez-Donaire}
\affiliation{Harvard-Smithsonian Center for Astrophysics, 60 Garden St., Cambridge, MA 02138, USA}

\author{Andreas Schruba}
\affiliation{Max-Planck-Institut f\"ur extraterrestrische Physik, Giessenbachstra\ss{}e 1, D-85748 Garching, Germany}

\author{Frank Bigiel}
\affiliation{Institut f\"ur theoretische Astrophysik, Zentrum f\"ur Astronomie der Universit\"at Heidelberg, Albert-Ueberle Str. 2, D-69120 Heidelberg, Germany}

\author{Molly Gallagher}
\affiliation{Department of Astronomy, The Ohio State University, 140 West 18th Avenue, Columbus, OH 43210, USA}

\author{Kelsey Johnson}
\affiliation{Department of Astronomy, University of Virginia, 520 McCormick Road, Charlottesville, VA 22904-4325, USA}

\author{Antonio Usero}
\affiliation{Observatorio Astron\'omico Nacional (IGN), C/ Alfonso XII 3, Madrid E-28014, Spain}

\begin{abstract}

Dense molecular gas and star formation are correlated in galaxies. The effect of low metallicity on this relationship is crucial for interpreting observations of high redshift galaxies, which have lower metallicities than galaxies today. However, it remains relatively unexplored because dense molecular gas tracers like \hcn\ and \hcop\ are faint in low metallicity systems. We present Green Bank Telescope observations of  \hcn(1-0) and \hcop(1-0)  on giant molecular cloud (34\,pc) scales in the nearby low metallicity ($12+\log({\rm O/H})=8.2$) starburst IC 10 and compare them to those in other galaxies. We detect \hcn\ and \hcop\ in one and three of five pointings, respectively. The \IHCNHCOP\ values are within the range seen in other galaxies, but are most similar to those seen in other low metallicity sources and in starbursts. The detections follow the fiducial \LIR-\LHCN\ and \LIR-\LHCOP\ relationships. These trends suggest that \hcn\ and \hcop\ can be used to trace dense molecular gas at metallicities of 1/4 $Z_\odot$, to first order. The dense gas fraction is similar to that in spiral galaxies, but lower than that in U/LIRGs. The dense molecular gas star formation efficiency, however, is on the upper end of those in normal galaxies and consistent with those in U/LIRGs. These results suggest that the CO and \hcn/\hcop\ emission occupy the same relative volumes as at higher metallicity, but that the entire emitting structure is reduced in size. Dense gas mass estimates for high redshift galaxies may need to be corrected for this effect.


\end{abstract}



\section{Introduction} \label{sec:introduction}

Within the Milky Way and in other nearby galaxies, dense molecular gas is closely linked to star formation \citep[e.g.,][]{Gao2004a,Gracia-Carpio2006IsGalaxies,Gracia-Carpio2008EvidenceGalaxies,Garcia-Burillo2012Star-formationGalaxies,Juneau2009,Kepley2014M82,Usero2015VariationsGalaxies}. Because of their high critical densities (${\sim}10^4-10^5 \, \rm{cm^{-3}}$), the fundamental rotational transitions of \hcn\ and \hcop\ are commonly used to trace this gas. Although these transitions are among the brightest available high critical density lines, they are still typically ${\sim}10-30$ times fainter than the brightest molecular gas tracer, \coone, in normal star-forming galaxies \citep[e.g.,][]{Usero2015VariationsGalaxies}. The faint nature of the emission makes them time-consuming to detect even in the brightest galaxies, limiting the range of galaxy types that have been explored thus far using these tracers. This picture is changing, however, now that the  100-m Green Bank Telescope (GBT) is able to regularly observe at millimeter wavelengths and the Atacama Large Millimeter/submillimeter Array (ALMA) is in full operations as well as the continuing success of the IRAM facilities. These instruments have the sensitivity to detect \hcn\ and \hcop\ across galaxies with a wide range of star formation rates, masses, and galaxy types \citep[e.g.,][]{Kepley2014M82,Usero2015VariationsGalaxies,Bigiel2016THEM51}. 

One area that remains relatively unexplored is the relationship between the dense molecular gas traced by \hcn\ and \hcop, star formation, and metallicity. Low metallicity dwarf galaxies, especially those that are gas rich and actively star-forming, occupy an important part of this parameter space. These local systems have similar metallicities to the mean metallicity observed in galaxies at $z\sim 2-3$ \citep{Yuan2013TheLensing}. Although they are not exact analogs to high-redshift galaxies, dwarf galaxies do provide key insights into how metallicity influences  dense gas tracers and their relationship to star formation, which are necessary for interpreting observations of dense gas at high redshifts.

Our understanding of how CO -- the primary tracer of the bulk molecular gas --  behaves at low metallicity suggests two possibilities for their behavior of \hcn\ and \hcop\ at low metallicity. CO emission from galaxies with metallicities less than solar is fainter than one would expect extrapolating from observations of higher metallicity sources \citep{Schruba2012,Taylor1998}. This result has been attributed to the reduced dust content of these systems, which leads to reduced shielding from dissociating radiation for CO \citep{Maloney1988,Wolfire2010}. The molecular hydrogen ($H_2$) is largely unaffected because it can shield itself from the dissociating radiation. This disparity drives the CO emission deeper into the molecular cloud as has been demonstrated observationally, e.g., using high (2pc) spatial resolution ALMA observations of CO in the low metallicity galaxy NGC6822 \citep{Schruba2017PhysicalWay}. 

If molecular clouds in low metallicity galaxies have the same density structure as at high metallicity, one might expect that while the CO emission is driven deeper into the cloud, the distribution of \hcn\ and \hcop\ emission is unchanged because it is tracing the dense molecular gas that remains well shielded from dissociating radiation. In that case we would expect the {\em apparent} $\fdense=\LHCNCO$, to be higher in low metallicity galaxies, while the dense gas star formation efficiency -- the ratio of star formation to \hcn\ (or \hcop\ emission)  -- would resemble that in normal star-forming galaxies. 

On the other hand, the whole emitting structure could be driven more deeply into the cloud. In this case, the radii of both the CO ($r_{CO}$) and HCN ($r_{HCN}$) emitting regions would be reduced but the ratio $r_{HCN}/r_{CO}$ would remain the same as at higher metallicity. Then we might expect the dense gas fraction  (i.e., \LHCNCO) in low metallicity sources to be similar to those found in higher metallicity galaxies because the relative size of the CO and HCN emitting regions is unchanged. The dense gas star formation efficiency, however, would be higher since the \hcn\ and \hcop\ emission is tracing a smaller part of the star-forming cloud. This toy model does not take into account that a significant amount of emission from a particular molecular species can be from gas with a density lower than its critical density \citep{Pety2017TheGalaxies,Kauffmann2017MolecularGalaxies}. However, as model calculations from \citet{Leroy2017Millimeter-WaveDistributions} have shown, the line ratios, which are the key observable above, still reflect the overall density distribution of gas.


In addition to the above density-based effects, the lower metallicities of dwarf galaxies may also change their molecular abundance patterns. Chemical modeling of low metallicity systems suggests that the relationship between metallicity and abundances of individual molecules is complex; a lower abundance of a particular element does not necessarily imply that all molecules containing that element have lower abundances \citep{Millar1990ChemicalSMC,Acharyya2015MOLECULARCLOUD,Acharyya2016SIMULATIONSCLOUD}. Comparing the properties of dense molecular gas tracers in a wide variety of systems as well as multiple dense gas tracers with different chemical pathways can yield insights into the effects of low metallicity on the abundances in the system.

Obtaining measurements of dense molecular gas tracers in low metallicity systems to test these hypotheses has been difficult.  Low metallicity galaxies already have faint CO emission \citep{Tacconi1987TheGalaxies,Taylor1998,Schruba2012,Cicone2017TheGalaxies} and even in solar metallicity sources the \hcn\ and \hcop\ are 10 to 30 times fainter than CO.  Observations of dense gas tracers in low metallicity galaxies have generally been limited to the four closest, actively star-forming dwarf irregulars --- the Magellanic Clouds \citep{Heikkila1999MolecularClouds,Chin1997MolecularCores,Chin1998MolecularCloud,Seale2012THECLOUD}, M33 \citep{Rosolowsky2011MinimalM33,Buchbender2013DenseHerM33es}, and NGC 6822 \citep{Israel2003Dust6822,Gratier2010The6822} -- and one more distant galaxy \citep[He 2-10;][]{Imanishi2009NobeyamaGalaxies,Santangelo2009Resolving2-10,Johnson2018The210}. The last galaxy is particularly notable, because He 2-10 is the only true dwarf starburst galaxy in this sample and thus bears the closest resemblance to sources at higher redshift.\footnote{We note that within He 2-10, the metallicity varies from super-solar in the star-forming clusters to values similar to the LMC ($12+\log{\rm (O/H)} = 8.3$) on the outer edges \citep{Cresci2017TheStarburst}.}

The observations of the Magellanic Clouds and the Local Group dwarf irregular NGC 6822 suggest that the dense gas fraction ($\propto \LHCNCO$) does appear higher in dwarfs than in more massive galaxies \citep{Chin1997MolecularCores,Chin1998MolecularCloud,Heikkila1999MolecularClouds,Israel2003Dust6822}. However, M33, a dwarf spiral with metallicity similar to the LMC, shows low ratios of dense gas tracers to CO emission \citep{Buchbender2013DenseHerM33es}.  Clearly, more data on dense molecular gas in low metallicity systems is needed to understand the impact of low metallicity on the dense gas tracers themselves and on the density structure of the molecular gas.

This paper uses observations of \hcn\ and \hcop\ emission from molecular clouds across the Local Group dwarf starburst IC 10 to explore how metallicity may affect the relationship between dense gas, CO emission, and star formation. Located only $817$~kpc away \citep{Sanna2008On10}, IC 10 has a metallicity of about $1/4$ $Z_\odot$ \citep[$12 + \log {\rm (O/H)} = 8.2 \pm 0.15$;][]{Lequeux1979ChemicalGalaxies,Garnett1990NitrogenGalaxies,Richer2001ICDwarf,Lee2003UncoveringField}, intermediate between that of the LMC and SMC. Recently, \hcn\ and \hcop\ emission have been detected in a single pointing in this galaxy using the Nobeyama Radio Observatory (NRO) 45\,m by \citet{Nishimura2016SpectralIC10} and in two out of three pointings using the IRAM 30\,m by \citet{Braine2017DenseGalaxies}. 

Here we present contemporaneous observations of five positions within IC 10 using the GBT 4mm receiver. These observations have ${\sim}2{-}3$ times higher resolution than those of \citet{Nishimura2016SpectralCloud} and \citet{Braine2017DenseGalaxies} allowing us to sample, for the first time, spatial scales similar to those of the giant molecular clouds in IC 10 (${\sim}34$~pc). The observed pointings in IC~10 were selected based on their high CO brightness and to sample different regions across the disk of IC~10. One of our pointings has also been observed by \citet{Nishimura2016SpectralIC10} and \citet{Braine2017DenseGalaxies} and two others by \citet{Braine2017DenseGalaxies} (see Figure~\ref{fig:spectra_overview}).  We combine our GBT data with maps of \coone\ emission, which traces the bulk of the molecular gas (modulo the caveats above), and far-infrared images, tracing recent star formation, to gain a more complete picture of the molecular gas and star formation in this galaxy. Finally, we compare our results with observations of dense molecular gas tracers in other systems including resolved observations of other low metallicity systems and of normal galaxies\footnote{Here we define normal galaxies as massive spiral galaxies on the star formation main sequence \citep[e.g.,][]{Saintonge2017XCOLDStudies}.} as well as integrated measurements of entire galaxies, luminous infrared galaxies (LIRGs), and ultra-luminuous infrared galaxies (ULIRGs).


\section{Data} \label{sec:data}

\subsection{GBT Data} \label{sec:gbt_data}

We used the 2-beam 4mm receiver on the Green Bank Telescope (GBT) to observe \hcn(1-0) ($\nu_{\rm rest} = 88.63160$ GHz) and \hcop(1-0) ($\nu_{\rm rest} = 89.18853$ GHz) at five different positions across the disk of IC~10 (GBT15A-159, PI: A.~Kepley).  The VEGAS backend was configured with two 187.5 MHz-wide spectral windows with 5.7kHz channels centered on the rest frequencies of the lines. The pointings within IC~10 were drawn from a list of CO 1-0 emission peaks identified using $^{12}$CO(1-0) and $^{12}$CO(2-1) single dish observations from the IRAM 30-m telescope (Bittle et al, in prep). At the distance of IC 10 \citep[$817$~kpc;][]{Sanna2008On10}, the $8.5\arcsec$ resolution of the GBT corresponds to ${\sim}34$~pc, approximately the size of an individual giant molecular cloud.

The detailed information for each observing session is given in Table~\ref{tab:obs_summary}. We began each observing session with out-of-focus holography scans to correct the dish surface. Each observing session included a set of nods taken on a flux calibrator. Pointing and focusing on calibration sources were done approximately every hour. After the pointing and focusing, we usually carried out a brief nod on the pointing and focus calibrator to serve as a check on telescope performance. We generally observed one pointing until we either had a detection or a strong limit on the line emission before moving on to the next pointing.

Observations of the source nodded the target between the two beams of the 4mm receiver. The data were taken in 20~minute blocks with 30s scans alternating between the two beams. Calibration wheel scans with a hot and cold load were done at the beginning and end of each block.

The likelihood of contamination in the off spectrum is small. The separation between the two beams of the 4mm receiver ${\sim}4.5\arcmin$, which is enough that the off-source beam lies away from the galaxy. In addition,  IC 10 has a  $v_{\rm LSR} \sim -350$ km~s$^{-1}$, so its emission is not blended with that from the Milky Way. 

We calibrated the data in GBTIDL \citep{Marganian2006GBTIDL:Data}\footnote{http://gbtidl.nrao.edu}. Because the signal paths for the two beams are separate, the data for each beam were calibrated independently. First, the calibration wheel scans were used to convert instrumental voltages to antenna temperatures. Then atmospheric corrections as a function of time were applied for each night based on weather data\footnote{http://www.gb.nrao.edu/∼rmaddale/Weather}. Next, main beam and aperture efficiencies were calculated for each night from the nod scans targeting a calibrator of known flux, with the current flux density of the calibrator taken from closest observation in frequency and time in the CARMA calfind database\footnote{https://carma.astro.umd.edu/cgi-bin/calfind.cgi}. The efficiencies we derive here are consistent with those in \citet{Kepley2014M82}. The exception to the above approach was the observing session on 2015 July 23. For this session, the calibrator scans were insufficient to calculate the efficiencies. Instead we used the average value of the main beam efficiency from the other nights. Finally, main beam efficiency was applied to the data, and the scans from both beams were averaged together and a DC baseline was subtracted to create a final spectrum on a ${T_{\rm MB}}$ scale with 1 \kms\ channels.


\begin{deluxetable*}{lccccc}
\tablecolumns{6}
\tablewidth{0pt}
\tabletypesize{\scriptsize}
\tablecaption{Observation Summary \label{tab:obs_summary}}
\tablehead{
\colhead{UT Date} &
\colhead{ 4 Feb 2015} &
\colhead{ 6 Feb 2015 } &
\colhead{14 May 2015} &
\colhead{24 May 2015} &
\colhead{ 24 Jul 2015} 
}
\startdata
UT Times & 1:00-4:45 & 1:30-5:15 & 9:00-12:00 & 5:45-11:00 & 3:30-10:30 \\
Flux Calibrator & 0319+4130, 0927+3902 & 0319+4130 & 2253+1608 & 2202+4216 & 1642+3948  \\
Flux Value (Jy) & 17.3, 3.8 & 17.3 & 18.0 & 3.8 & 2.1\tablenotemark{b} \\
Average Opacity & 0.0829 & 0.0577 & 0.1343 & 0.1948 & 0.4169 \\
Average $T_{\rm sys}$ \tablenotemark{a} (\hcn, \hcop) (K) & 133, 65 & 125, 69 & 130, 121 & 156, 125 & 120, 115 \\
Main Beam Efficiency (\hcn, \hcop) & 0.304, 0.301 & 0.194, 0.200 & 0.280, 0.270 & 0.263, 0.255 & 0.260, 0.257\tablenotemark{b} \\
\enddata
\tablenotetext{a}{Not corrected for atmospheric contribution.}
\tablenotetext{b}{The flux calibrator data for this night was
  insufficient for calculating the main beam efficiency. We
  adopted average efficiencies from previous nights for this night. }
\end{deluxetable*}

\subsection{CO Data} \label{sec:CO_data}

We use \co{12}{1}{0} maps from CARMA and the IRAM 30-m telescope to trace the bulk molecular gas content (Bittle et al, in prep). These data are vastly more sensitive than the previous BIMA \co{12}{1}{0} map by \citet{Leroy2006Molecular10} and include short spacing information. The molecular peaks we target are visible in the earlier paper and are consistent with the new data. The combined IRAM 30-m and CARMA data have a synthesized beam $\sim 8\arcsec$, 2.5~\kms\ channels, an rms of 60~mK, and are sensitive to emission on large scales.

\subsection{IR Data} \label{sec:IR_data}

We calculate \LIR , defined here as total infrared luminosity integrated over the entire infrared band (8-1000\micron), for each GBT pointing using data from the PACS instrument on {\it Herschel}. To do this, we first retrieved the standard level~2 products for IC 10 from the Herschel archive. Ideally, one would use the 70\micron, 100\micron, and 160\micron\ data to derive \LIR\ for each GBT pointing. This wavelength range contains 30-50\% of the IR emission from a source and has a lower scatter in derived \LIR\ values than just using 70\micron\ and 100\micron\ \citep{Galametz2013CalibrationBands}. However, the resolution of the 160\micron\ Herschel image is coarser than the GBT beam (11.4\arcsec\ vs. 8.5\arcsec). To work around this, we smooth the 70\micron, 100\micron, and 160\micron\ images to a common 15\arcsec\ Gaussian beam using the method in \citet{Aniano2011Common-ResolutionTelescopes}. Then we use the coefficients given in line 17 of Table 3  of \citet{Galametz2013CalibrationBands} to calculate the total IR surface brightness, \SIR, from the combination of the smoothed 70\micron, 100\micron, and 160\micron\ data. From this, we calculate the ratio of 70\micron\ emission to \SIR\ for each point. These ratios ranged from 1.3 for point F to 3.5 for point E. Next we convolve the {\em Herschel} 70\micron\ data to match the beam of the GBT data (8.5\arcsec) and divide by the 70\micron\ to \SIR\ ratio calculated above to estimate the \SIR\ value for each point. Finally, we converted this 8.5\arcsec\ \SIR\ value to an \LIR\ value using our adopted distance. The derived \LIR\ values are given in Table~\ref{tab:properties}. 

\section{Results} \label{sec:results}

\subsection{\hcn\ and \hcop\ Detections and Limits} \label{sec:detections}

\begin{figure*}
\centering
\includegraphics[height=0.8\textheight]{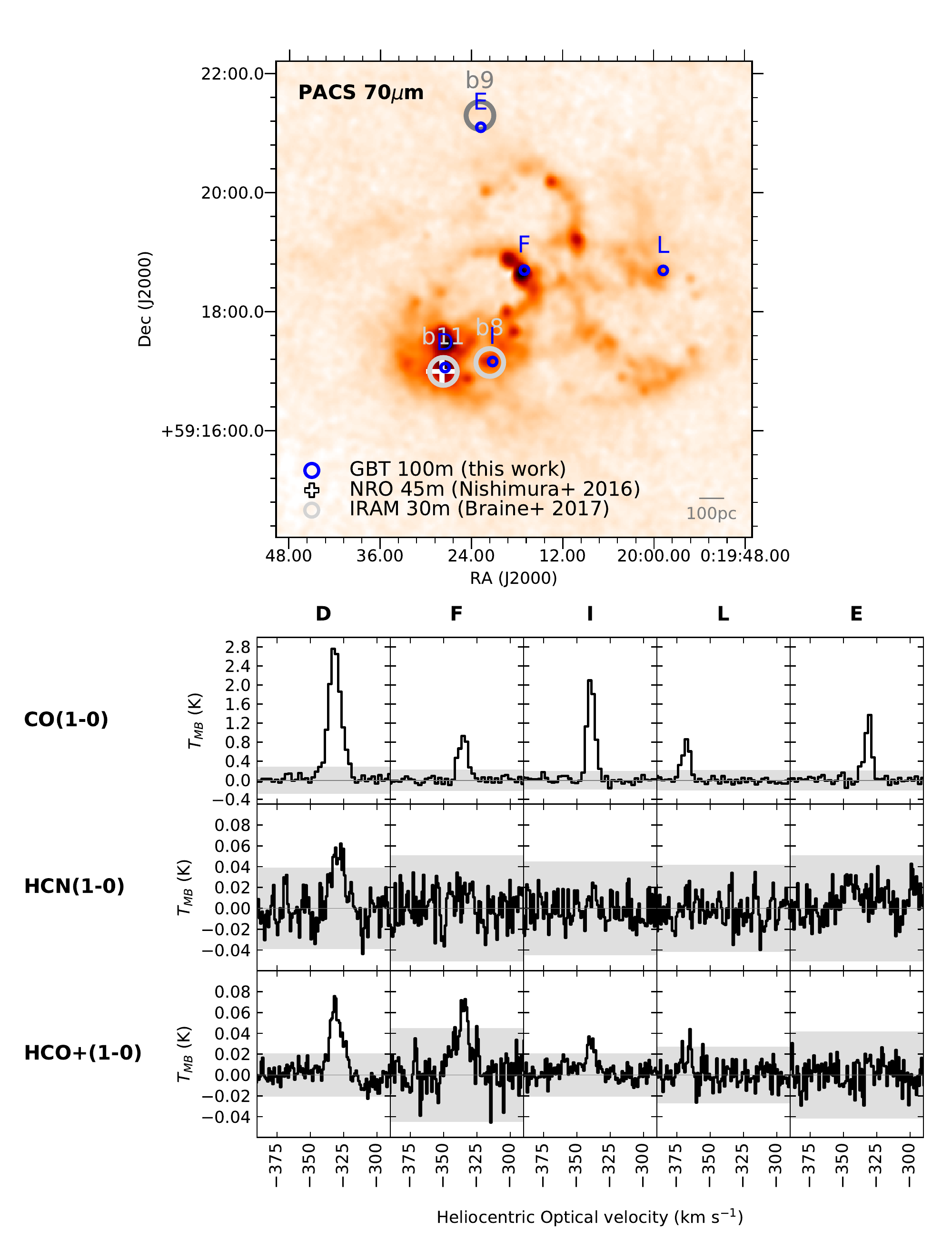}
\caption{In the low metallicity galaxy IC 10, we have detected \hcn\ emission in one out of the five observed positions  and \hcop\ emission in three of the five observed positions  with a tentative detection of \hcop\ at position L. {\em Top:} PACS 70\micron\ image with positions observed with the GBT shown as blue circles. The position of the \hcn\ detection by \citet{Nishimura2016SpectralIC10} is shown as a white cross. The pointings observed by \citet{Braine2017DenseGalaxies} are shown as gray circles and are labeled with the identifier from that work. The size of the symbols for each data set corresponds to the beam for each set of observations: 8.5\arcsec\ (GBT), 20.4\arcsec\ (NRO), 28\arcsec\ (IRAM). {\em Bottom:} Spectra of \coone\ (top row), \hcn\ (middle row), and \hcop\ (bottom row) from each GBT pointing. The systemic velocity of IC 10 is -348~\kms\ using the optical convention in the heliocentric frame \citep{Huchra1999TheCap}. The $\pm 3\sigma$ region is indicated as a shaded gray region about the zero baseline shown as a dark gray line.}
\label{fig:spectra_overview}
\end{figure*}

Figure~\ref{fig:spectra_overview} shows the locations of our five observed pointings on the PACS 70$\mu$m map, along with the CO, \hcn, and \hcop\ spectra for each pointing. We detect \hcop\ in three of the five surveyed positions (D, F, and I) with a tentative detection at one position (L). We detect \hcn\ in one position (D). We identify detections based on 1) the presence of  emission greater than 3$\sigma$ within $\pm$ 250 \kms\ of the LSR velocity of CO and 2) that the fitted peak velocity of this emission is within $\pm5 \ \kms$ of the CO. Requiring both conditions 1 and 2 allows us to assess the possibility of false positives. We do not find any $>3\sigma$ peaks for \hcn\ and \hcop\ more than 3\kms\ away from the derived \coone\ peak.  Our tentative \hcop\ detection at position L is based on the presence of significant ($>3\sigma$) emission at the same velocity as the \coone.  The properties of the line emission were derived by fitting the spectra with a single Gaussian. For non-detections, 3$\sigma$ upper limits on the peak $T_{\rm MB}$ were derived from the standard deviation of the spectrum and 3$\sigma$ upper limits on the luminosity were derived assuming a line width of $10$~\kms\ and unresolved sources. These results are summarized in Table~\ref{tab:properties}.

Some of our observed positions were also observed at lower resolution by \citet{Nishimura2016SpectralIC10} and \citet{Braine2017DenseGalaxies}. The GBT observations presented here sample 34~pc size scales, while the \citet{Nishimura2016SpectralIC10} observations sample 80~pc size scales and the \citet{Braine2017DenseGalaxies} observations sample 110~pc size scales. Our position~D is coincident with the  position observed by \citet{Nishimura2016SpectralIC10}. The \hcn\ and \hcop\ linewidths and velocities presented here are consistent with the detections in that paper, although the \citeauthor{Nishimura2016SpectralIC10} detection has a slightly larger linewidth (${\sim} 14{-}15~\kms$ vs. ${\sim} 11~\kms$). The larger line width measured by \citeauthor{Nishimura2016SpectralIC10} may result from their larger beam size (20\arcsec\ vs. 8.5\arcsec) sampling a wider range of velocities, which would be expected when observing a turbulent medium at larger scales. Positions b8, b9, and b11 observed in \citet{Braine2017DenseGalaxies} are coincident with our positions I, E, and D, respectively.  The latter paper does not measure the peaks and linewidths, but a comparison of the detections shown in their Figure~5 with our detections show similar profiles, but with lower line peaks and broader widths.

We find that the measured integrated line luminosity for position D -- the only position detected in  \citet{Nishimura2016SpectralIC10}, \citet{Braine2017DenseGalaxies}, and this paper -- increases with increasing beam size.  Compared to our derived line luminosity, the measured \hcn\ line luminosities for position D are 2.5 times and 3.7 times higher  and the measured \hcop\ line luminosities are 5.7 times and 9.8 times higher for \citet{Nishimura2016SpectralIC10} and \citet{Braine2017DenseGalaxies}, respectively. Here we assume that differences in calibration and data processing among the data sets are negligible. This trend suggests that the \hcn\ and \hcop\ emission is spread across a region at least $\sim 110 {\rm pc}$, i.e., the  \citet{Braine2017DenseGalaxies} beam size. This result is supported by the larger line widths found by \citeauthor{Nishimura2016SpectralIC10} and \citeauthor{Braine2017DenseGalaxies} In addition, both \citet{Nishimura2016SpectralIC10} and \citet{Braine2017DenseGalaxies} have more excess \hcop\ emission  than they have excess \hcn\ emission. This trend also suggests that the \hcop\ emission is more extended than the \hcn\ emission. 

\begin{deluxetable*}{cllllcrrrr}
\tablewidth{0pt}
\tabletypesize{\scriptsize}
\tablecaption{Properties of Detected Regions \label{tab:properties}}
\tablecolumns{10}
\tablehead{
    \colhead{} & 
    \colhead{R.A.} & 
    \colhead{Decl.} & 
    \colhead{\LIR} & 
    \colhead{} & 
    \colhead{$\sigma$} & 
    \colhead{Peak $T_{MB}$\tablenotemark{b}} & 
    \colhead{FWHM} & 
    \colhead{$V_0$\tablenotemark{c}} & 
    \colhead{Luminosity\tablenotemark{d}} \\ 
    \colhead{Region\tablenotemark{a}} & 
    \colhead{(J2000)} & 
    \colhead{(J2000)} & 
    \colhead{($10^6 \Lsun$)} & 
    \colhead{Line} & 
    \colhead{(K)} & 
    \colhead{(K)} & 
    \colhead{(\kms)} & 
    \colhead{(\kms)} & 
    \colhead{($10^3 \Kkmspc$)}  
    }
\startdata
D & 00:20:27.480 & +59:17:03.84 & $   4.13$ & ${\rm HCN (1-0)}$ & $  0.013$ & $  0.054 \pm   0.006$ &  $   11.5 \pm     1.3$ &  $ -329.4 \pm     0.6$  &  $   0.92 \pm    0.08$  \\ 
 &  &  &  & ${\rm HCO^+ (1-0)}$ & $  0.007$ & $  0.064 \pm   0.003$ &  $   11.0 \pm     0.6$ &  $ -330.8 \pm     0.3$  &  $   1.02 \pm    0.05$  \\ 
 &  &  &  & \coone & $  0.095$ & $   2.74 \pm    0.06$ &  $   11.2 \pm     0.3$ &  $ -331.6 \pm     0.1$  &  $  43.23 \pm    0.67$  \\ \hline 
E & 00:20:22.793 & +59:21:06.12 & $   0.06$ & ${\rm HCN (1-0)}$ & $  0.017$ & $    <   0.052$ &  \nodata &  \nodata  &  $    <    0.34$  \\ 
 &  &  &  & ${\rm HCO^+ (1-0)}$ & $  0.014$ & $    <   0.041$ &  \nodata &  \nodata  &  $    <    0.29$  \\ 
 &  &  &  & \coone & $  0.069$ & $   1.36 \pm    0.06$ &  $    5.8 \pm     0.3$ &  $ -331.2 \pm     0.1$  &  $  11.54 \pm    0.56$  \\ \hline 
F & 00:20:17.038 & +59:18:42.12 & $   1.89$ & ${\rm HCN (1-0)}$ & $  0.017$ & $    <   0.050$ &  \nodata &  \nodata  &  $    <    0.35$  \\ 
 &  &  &  & ${\rm HCO^+ (1-0)}$ & $  0.015$ & $  0.056 \pm   0.006$ &  $   14.8 \pm     1.7$ &  $ -335.6 \pm     0.7$  &  $   1.23 \pm    0.11$  \\ 
 &  &  &  & \coone & $  0.075$ & $   0.94 \pm    0.06$ &  $    8.4 \pm     0.6$ &  $ -335.3 \pm     0.2$  &  $  10.89 \pm    0.52$  \\ \hline 
I & 00:20:21.216 & +59:17:09.96 & $   0.58$ & ${\rm HCN (1-0)}$ & $  0.015$ & $    <   0.045$ &  \nodata &  \nodata  &  $    <    0.30$  \\ 
 &  &  &  & ${\rm HCO^+ (1-0)}$ & $  0.007$ & $  0.032 \pm   0.004$ &  $    8.5 \pm     1.1$ &  $ -339.6 \pm     0.5$  &  $   0.43 \pm    0.05$  \\ 
 &  &  &  & \coone & $  0.066$ & $   2.17 \pm    0.06$ &  $    6.9 \pm     0.2$ &  $ -339.4 \pm     0.1$  &  $  20.29 \pm    0.55$  \\ \hline 
L & 00:19:58.752 & +59:18:41.76 & $   0.25$ & ${\rm HCN (1-0)}$ & $  0.014$ & $    <   0.042$ &  \nodata &  \nodata  &  $    <    0.31$  \\ 
 &  &  &  & ${\rm HCO^+ (1-0)}$ & $  0.009$ & $  0.041 \pm   0.008$ &  $    2.8 \pm     0.6$ &  $ -365.3 \pm     0.2$  &  $   0.16 \pm    0.04$  \\ 
 &  &  &  & \coone & $  0.072$ & $   0.84 \pm    0.06$ &  $    7.0 \pm     0.6$ &  $ -368.1 \pm     0.2$  &  $   8.11 \pm    0.45$  \\  
\enddata 
\tablenotetext{a}{The observed pointings were drawn from a longer list of \co{12}{1}{0} peaks (Bittle et al., in prep) and thus do not start with A and are not consecutive.}
\tablenotetext{b}{3$\sigma$ upper limits derived from the standard deviation of the spectrum. If a peak was found, then the error from a fit to a Gaussian is given.}
\tablenotetext{c}{LSR radio velocity}
\tablenotetext{d}{3$\sigma$ upper limits derived from the standard deviation of the spectrum and assuming a line width of 10\kms and unresolved emission}
\end{deluxetable*}

\subsection{The Effect of Metallicity on \IHCNHCOP} \label{sec:hcn_hcop}

\begin{figure*}
\centering
\includegraphics[width=0.7\textwidth]{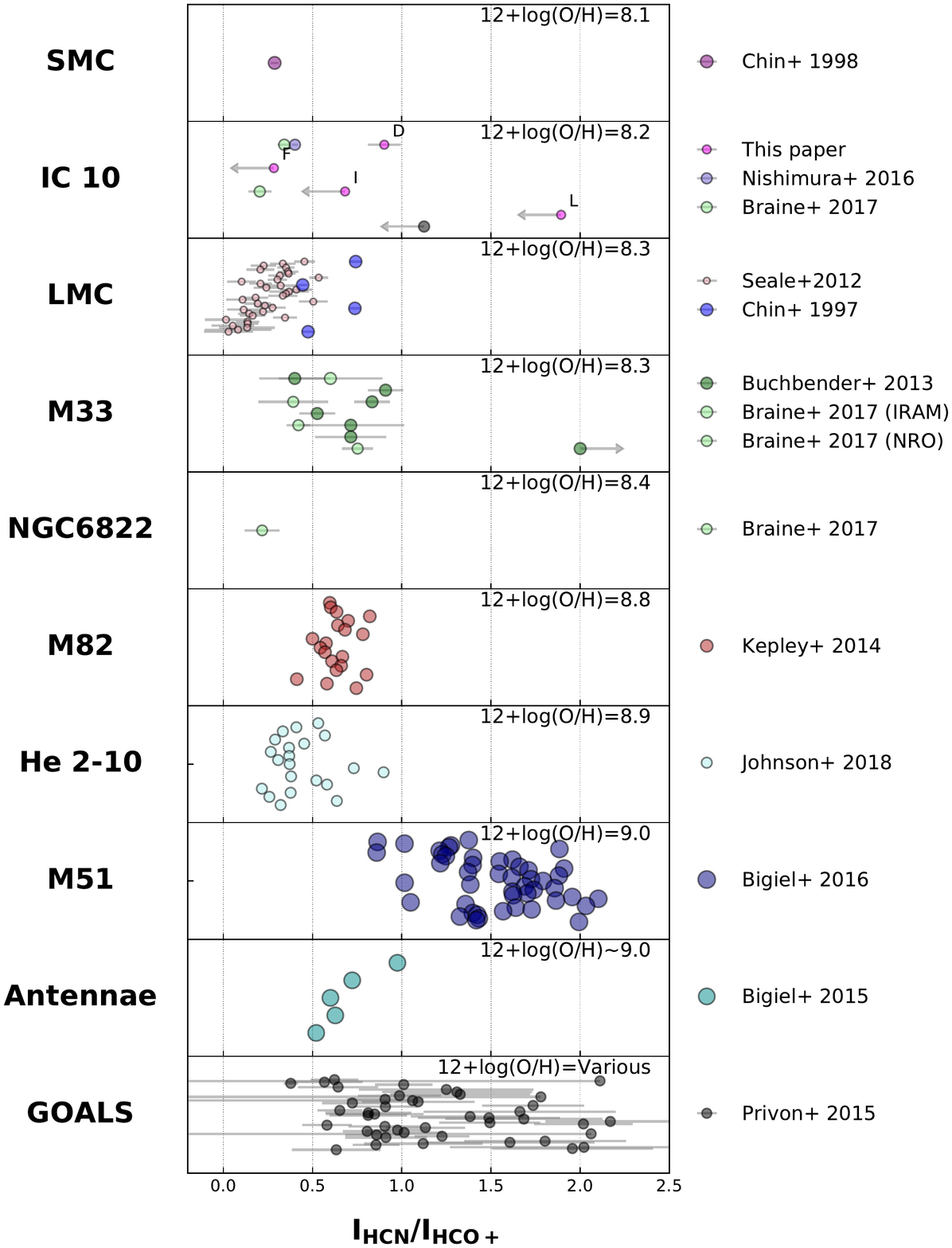}
\caption{\hcn/\hcop\ ratio for several low metallicity systems (SMC, IC10, LMC, M33, NGC6822), a ``typical" spiral (M51), a starburst galaxy (M82), a dwarf starburst (He 2-10), a late stage merger (the Antennae), and a large sample of infrared bright galaxies (GOALS). The literature values are taken from \citet{Nishimura2016SpectralIC10}, \citet{Chin1997MolecularCores}, \citet{Chin1998MolecularCloud}, \citet{Buchbender2013DenseHerM33es}, \citet{Braine2017DenseGalaxies}, \citet{Bigiel2016THEM51}, \citet{Kepley2014M82}, \citet{Bigiel2015DENSEGALAXIES}, \citet{Johnson2018The210}, and  \citet{Privon2015EXCITATIONSURVEY}. The metallicity  for each galaxy is shown in the upper right hand corner and the placement of the points on the vertical axis for each source is arbitrary except for IC 10, where we have placed measurements from the same location in the galaxy at the same (arbitrary) y value. We use metallicities taken from the compilation of \citet{Krumholz2011} rather than individual metallicity estimates from the literature because that work placed all the galaxies on the same metallicity scale. The size of points in the plot have been scaled logarithmically with the spatial resolution of the observation, except for the points from \citet{Privon2015EXCITATIONSURVEY}, which span a wide range of spatial scales. In general, the \hcn/\hcop\ ratios in IC~10 are less than one. They are within the range of values seen in other higher metallicity sources. They are most similar to the ratios seen in other low metallicity galaxies and in starburst galaxies like M82 and He 2-10. Our point D is coincident with the \citet{Nishimura2016SpectralIC10} detection and pointing b11 from \citet{Braine2017DenseGalaxies}, but we measure a higher \hcn/\hcop\ intensity ratio than either. This is likely the result of our smaller beam size and, taken at face value, suggests that the \hcn\ emission is distributed over a smaller area than the \hcop\ emission.}
\label{fig:hcn_hcop}
\end{figure*}

To investigate the potential effects of low metallicity on the \hcn\ and \hcop\ measurements, Figure~\ref{fig:hcn_hcop} compares the measured \IHCNHCOP\ line ratios (including upper limits) with those found in other low metallicity systems \citep{Chin1997MolecularCores,Chin1998MolecularCloud,Buchbender2013DenseHerM33es,Braine2017DenseGalaxies}, the normal spiral M51 \citep{Bigiel2016THEM51}, and in other starbursting galaxies like M82 \citep{Kepley2014M82}, the Antennae \citep{Bigiel2015DENSEGALAXIES}, and He 2-10 \citep{Johnson2018The210}.

In three of the four IC 10 pointings with at least one line detected (D, F, and I), the \IHCNHCOP\ ratio is constrained to be less than one. The remaining pointing (L) has a tentative \hcop\ detection and an upper limit on \IHCNHCOP\ that is consistent with unity.  The \IHCNHCOP\ ratio for position~D is higher than that found at the same position by \citet{Nishimura2016SpectralIC10} and \citet{Braine2017DenseGalaxies}. To show this, we have given all three data points the same (arbitrary) y-axis value in Figure~\ref{fig:hcn_hcop}. The simplest explanation of this difference is the larger beam size of the latter observations. Resolved observations of \hcn\ and \hcop\ in low metallicity systems have found that the \hcop\ emission is typically more extended than the \hcn\ emission. This effect has been seen in observations on both large (${\sim} 150$~pc; \citealp{Johansson1994InterstellarCloud,Heikkila1999MolecularClouds}) and small (${\sim} 15$~pc); \citealt{Seale2012THECLOUD,Pety2017TheGalaxies}) scales. Therefore, the larger (${\sim} 80$~pc) beam of \citeauthor{Nishimura2016SpectralIC10} could be expected to capture more \hcop\ emission than our smaller ${\sim} 34$~pc beam, resulting in a lower \IHCNHCOP\ ratio. Mapping, e.g., with the new GBT 16-pixel ARGUS array, would conclusively test this hypothesis. 

Figure~\ref{fig:hcn_hcop} shows that these measured \IHCNHCOP\ line ratios (including upper limits) in IC 10 are within the range of values found for all our comparison samples. They are most consistent with those found in other low metallicity systems as well as those in other starbursting galaxies like M82, He 2-10, and the Antennae \citep{Kepley2014M82,Bigiel2015DENSEGALAXIES,Johnson2018The210}. They are on the low end of the \IHCNHCOP\ ratios found for the GOALS sample of bright infrared galaxies \citep{Privon2015EXCITATIONSURVEY} and lower than the ratios found in the nearby spiral galaxy M51  \citep{Bigiel2016THEM51,Gallagher2018DenseGalaxies}. The line ratios found in the latter system are consistent with the line ratios found for a larger sample of 9 spiral galaxies (M. Jim\'enez-Donaire et al., in prep). Aside from the differences in \IHCNHCOP\ seen for position D~in IC~10,  the \IHCNHCOP\ does not appear to depend strongly on the spatial resolution of the observations when compared among different sources (see Figure~\ref{fig:hcn_hcop}).

The observed values of \IHCNHCOP\ in IC 10 could potentially be the result of a number of effects including  differences in the density distribution \citep{Leroy2017Millimeter-WaveDistributions} or abundance ratios  \citep{Nishimura2016SpectralCloud,Nishimura2016SpectralIC10} of the molecular gas as well as the effects of photo dominated regions \citep{Heikkila1999MolecularClouds} or shocks \citep{Mitchell1983EffectsCloud}. We do not have enough information here to distinguish between these scenarios.  Regardless of the exact mechanisms involved, however, the values of \IHCNHCOP\ in IC~10 and other low metallicity galaxies are similar to the low end of the values found in higher metallicity galaxies. This result suggests that changes in \IHCNHCOP\ due to metallicity are relatively small compared to the overall range of \IHCNHCOP\ values found within and among galaxies.

\subsection{The Effect of Metallicity on the Relationship Between Dense Gas and Star Formation} \label{sec:LIR-LHCN}

\begin{figure*}
\gridline{\fig{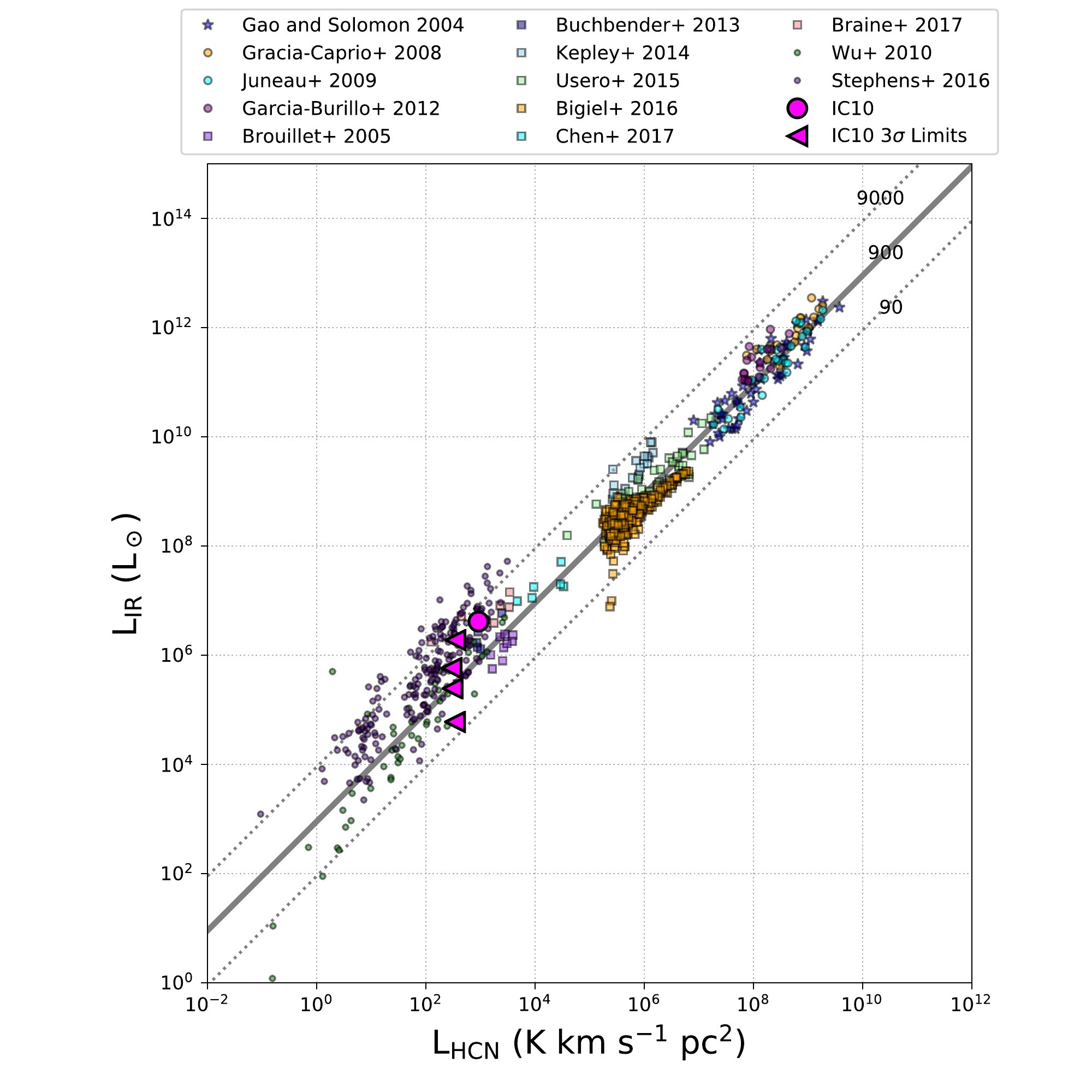}{0.43\textwidth}{(a)}
          \fig{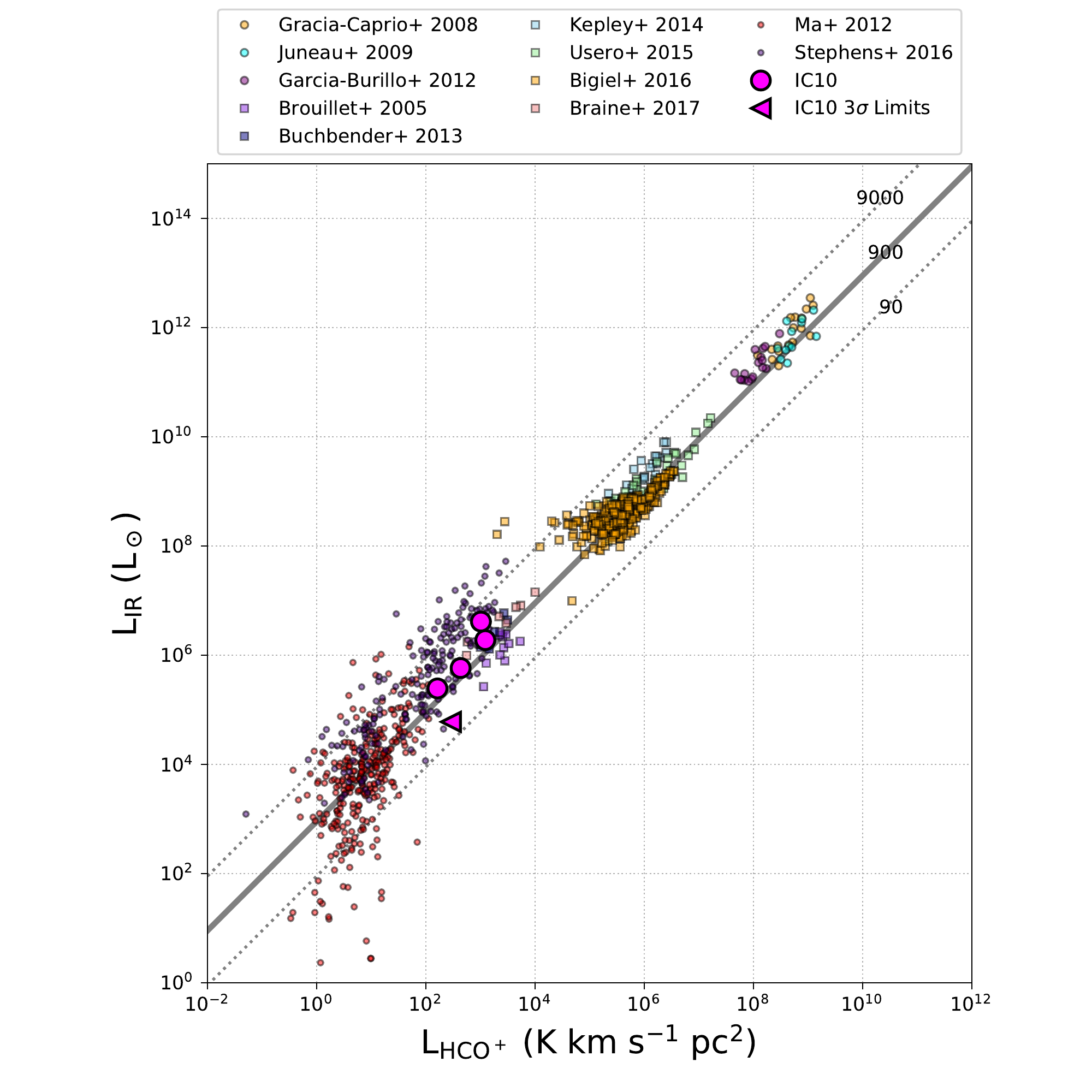}{0.43\textwidth}{(b)}}
\caption{\LIR\ as a function of luminosity of the dense gas tracers \hcn\ (left) and \hcop\ (right) compared with measurements from the literature including  integrated measurements of entire galaxies \citep[stars;][]{Gao2004}, integrated measurements of LIRGs and ULIRGs \citep[circles;][]{Gracia-Carpio2008EvidenceGalaxies,Juneau2009,Garcia-Burillo2012Star-formationGalaxies}, resolved measurements within galaxies \citep[squares;][]{Buchbender2013DenseHerM33es,Kepley2014M82,Usero2015VariationsGalaxies,Bigiel2016THEM51,Chen2017DenseM51}, 
and finally measurements for individual clouds in the Milky Way \citep[pentagons;][]{Wu2010TheCs,Ma2013TheClumps,Stephens2016LinkingGalaxies}. The fiducial \citet{Gao2004} fit ($\LIR/\LHCN=900\ \Lsun(\Kkmspc)^{-1}$) is shown as a dark gray line. We show the same line in the $\LIR{-}\LHCOP$ plot to facilitate comparison between the \hcn\ and \hcop\ plots. The dotted gray lines are 0.1 and 10 times the fiducial \citet{Gao2004a} relationship showing the large range of values for this relationship. The \hcn\ and \hcop\ detections and limits within IC 10 are consistent with fiducial relationships seen in e.g., \citet{Gao2004a} and \citet{Gracia-Carpio2008EvidenceGalaxies}, suggesting that to first order \hcn\ and \hcop\ are acceptable tracers of dense gas even in low metallicity systems like IC 10.} \label{fig:gao_and_solomon}
\end{figure*}

In the Milky Way and in nearby galaxies, the amount of dense molecular gas (as traced by \hcn\ or \hcop) is linearly correlated with the amount of star formation as traced by \LIR\ \citep[e.g.,][]{Gao2004a,Juneau2009,Wu2010TheCs}. To see whether the low metallicity of IC 10 affects this fiducial relationship, we plot the measured \LHCN\ and \LHCOP\ values as a function of \LIR\ for IC10 and compare them with other samples from the literature (Figure~\ref{fig:gao_and_solomon}). We include integrated measurements of entire galaxies \citep{Gao2004}, integrated measurements of LIRGs and ULIRGs \citep{Gracia-Carpio2008EvidenceGalaxies,Juneau2009,Garcia-Burillo2012Star-formationGalaxies}, resolved measurements within galaxies \citep{Brouillet2005HCNM31,Buchbender2013DenseHerM33es,Kepley2014M82,Usero2015VariationsGalaxies,Bigiel2016THEM51,Chen2017DenseM51}, and finally measurements for individual clouds in the Milky Way \citep{Wu2010TheCs,Ma2013TheClumps,Stephens2016LinkingGalaxies}. 

The \hcn\ and \hcop\ detections and limits that we measure in IC 10 are consistent with the correlation between IR emission and \hcn\ and \hcop\ first noted by \citet{Gao2004a}.  This result suggests that the low metallicity of IC 10  has a relatively small effect on \LHCN\ and \LHCOP. Combined with the knowledge that \IHCNHCOP\ also has a weak dependence on metallicity, our observations support the idea that \hcn\ and \hcop\ can be used as dense gas tracers in low metallicity galaxies like IC 10, at least to first order. However, we note that while the relationship space appears relatively tight  in log-log space ($\sigma_{\log \LIRHCN} \sim 0.3$), the \LIRHCN\ values have a large range (factor of 1000; cf. Figure~\ref{fig:gao_and_solomon}). We return to this point in \S~\ref{sec:dense_frac_SFE}.

It is important to note that using \LIR\ as a star formation indicator for IC 10 may introduce biases into this relationship. In general, extragalactic studies have focused on spiral galaxies and on dust-enshrouded starbursts like LIRGs and ULIRGs. Therefore, they have generally used \LIR\ as their star formation rate tracer of choice, since all or nearly all the photons emitted by young massive stars in these galaxies should have been reprocessed by dust.  For galaxies like IC 10, however, \LIR\ may underestimate the star formation rate in these systems due to their relative lack of dust, which allows photons from the young massive stars to escape without being processed by the dust \citep{Bell2003EstimatingCorrelation}.  

For IC 10, the integrated \ha-based star formation rate is a factor of 4 times higher than the integrated IR-based star formation rate \citep{Leroy2006Molecular10}. However, the observed pointings in IC 10 are bright in the IR (cf. Figure \ref{fig:spectra_overview}) and the GBT beam is roughly the same scale as the giant molecular clouds in this galaxy.  Therefore, we are likely probing the giant molecular cloud environment within IC 10, which should be relatively dusty, and therefore \LIR\ may be an appropriate tracer of star formation. High resolution observations of II~Zw~40, another low metallicity dwarf starburst, show significant extinction associated with massive star-forming regions even in this low metallicity system \citep{Kepley2014IIZw40}.  Based on the preceding arguments and for consistency with other observations, we use the measured \LIR\ as our star formation tracer for the individual pointings within IC 10, but note that the true star formation rates could potentially be a factor of ${\lesssim}\,4$ higher, i.e., the ratio of the \ha\ to \LIR-based star formation rate estimate for IC 10. If our assumed star formation rates for IC 10 are underestimated by up to a factor of 4, they would lie slightly above the fiducial relationship.

\subsection{The Relationship between the Dense Gas Fraction and the Molecular and Dense Gas Star Formation Efficiencies at Low Metallicity} \label{sec:dense_frac_SFE}

\begin{figure*}
\gridline{\fig{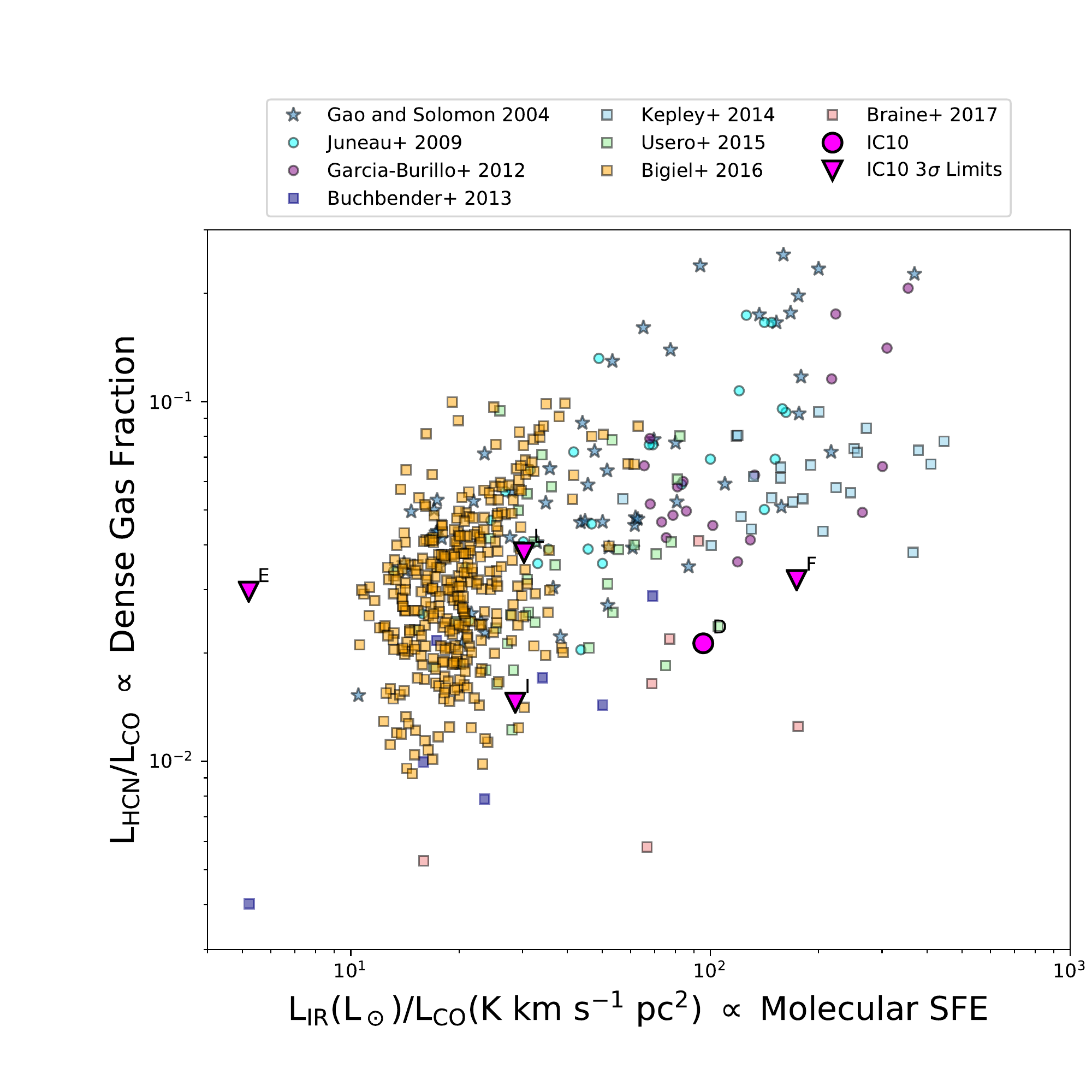}{0.43\textwidth}{(a)}
          \fig{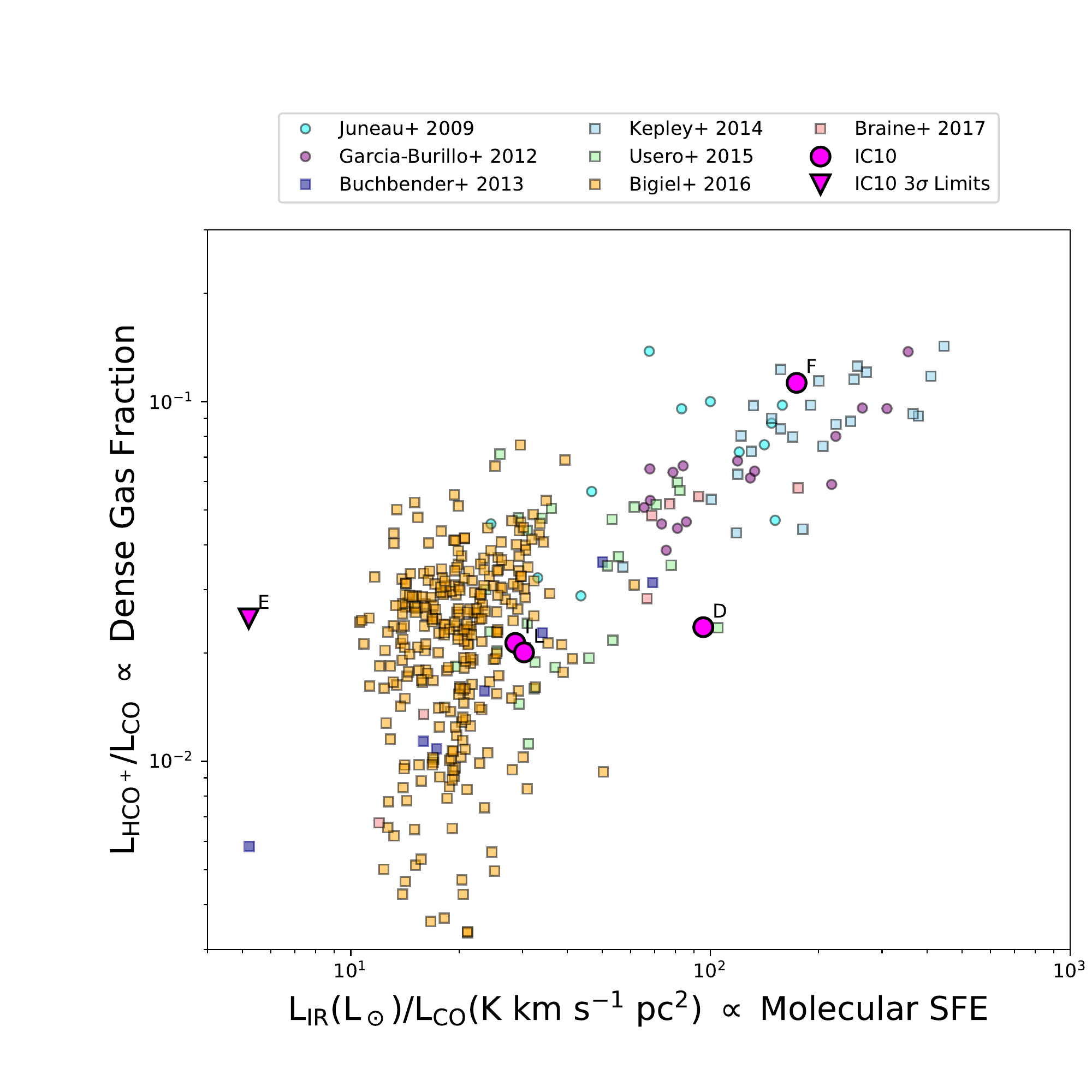}{0.43\textwidth}{(b)}}
\vspace*{-3mm}
\gridline{\fig{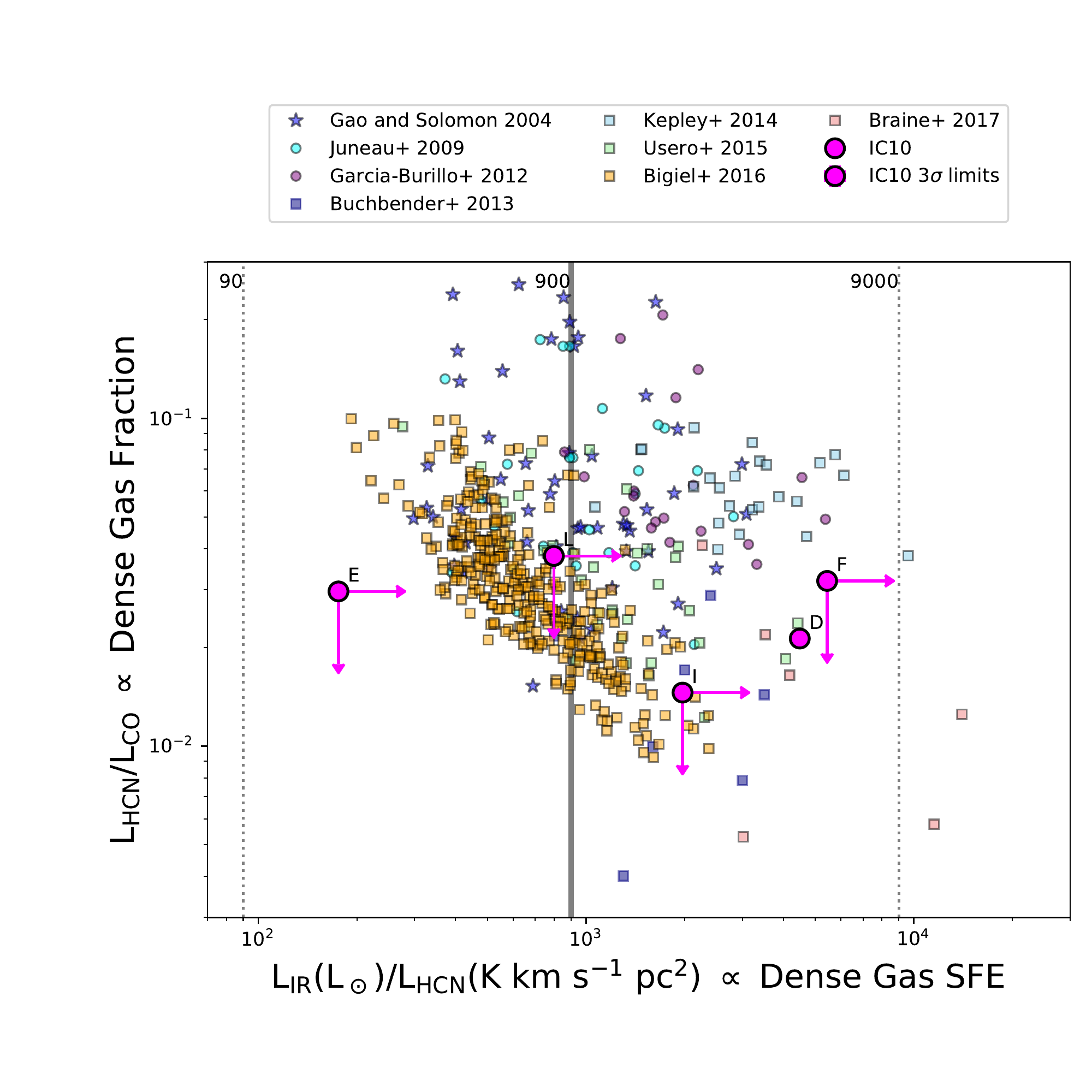}{0.43\textwidth}{(c)}
          \fig{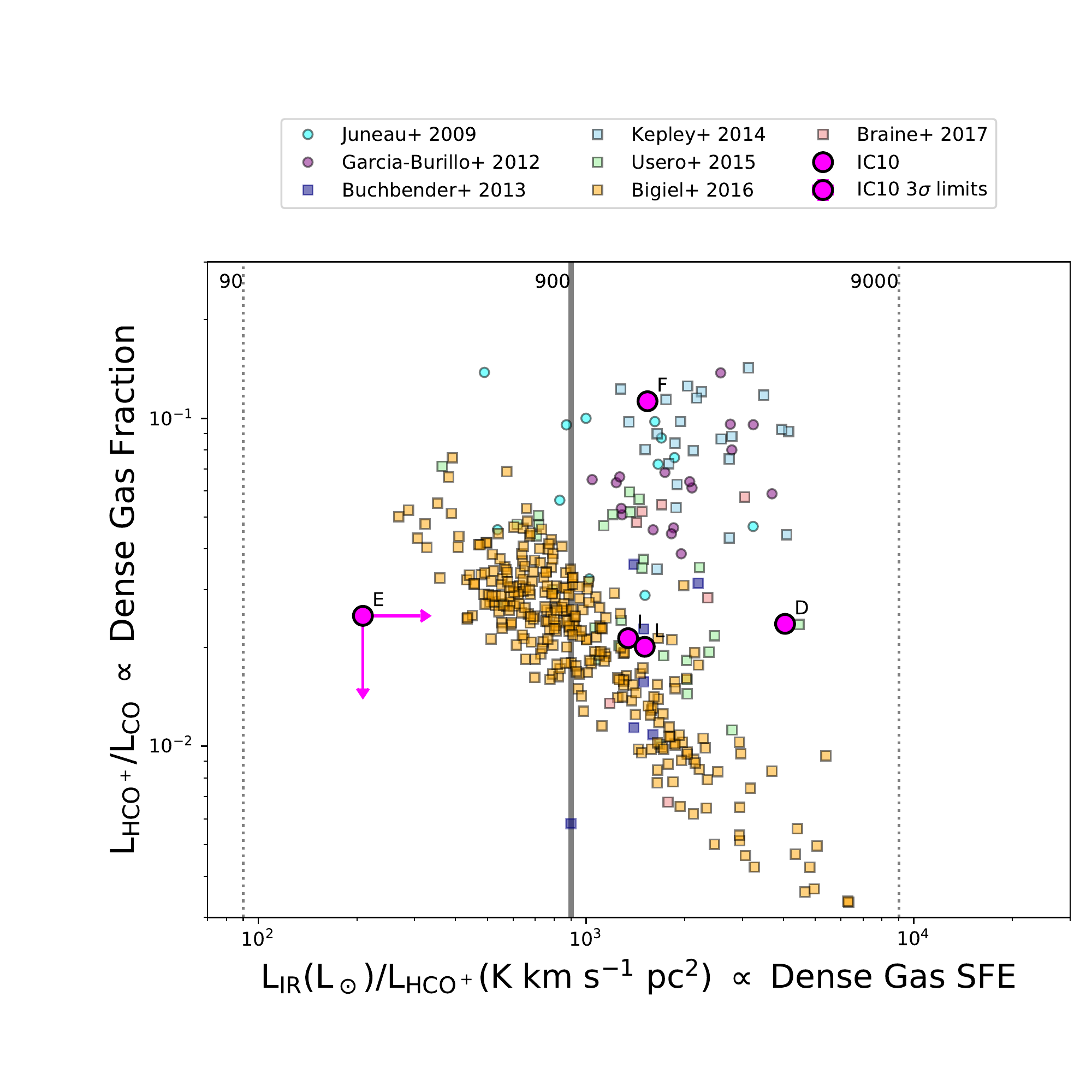}{0.43\textwidth}{(d)}}
\vspace*{-3mm}
\caption{{\em Top row:} Dense gas fraction as traced by \LHCNCO\ (left) and \LHCOPCO\ (right) as a function of molecular star formation efficiency (SFE) as traced by \LIRCO\ compared with measurements from the literature including  integrated measurements of entire galaxies \citep[stars;][]{Gao2004}, integrated measurements of LIRGs and ULIRGs \citep[circles;][]{Gracia-Carpio2008EvidenceGalaxies,Juneau2009,Garcia-Burillo2012Star-formationGalaxies}, and resolved measurements within galaxies \citep[squares;][]{Buchbender2013DenseHerM33es,Kepley2014M82,Usero2015VariationsGalaxies,Bigiel2016THEM51,Braine2017DenseGalaxies}. The molecular star formation efficiency appears to span the full range of molecular star formation efficiencies found in a sample of nearby galaxies, but is lower than that found in LIRGs/ULIRGS. With the exception of point~F, the dense gas fractions found in IC 10 are consistent with the average dense gas fractions found in individual regions within nearby galaxies by \citet{Usero2015VariationsGalaxies}. Point F, which is located at the intersection of several bubbles, has an elevated \LHCOPCO\ ratio that is likely the result of enhanced \hcop\ emission due to shocks.   {\em Bottom row:} The dense gas fraction as a function of the dense gas star formation efficiency for \hcn\ (left) and \hcop\ (right) for the same set of comparison samples. The vertical lines indicate from left to right 0.1 (dotted), 1 (solid), and 100 (dotted) times the fidicual \citet{Gao2004a} relationship. The points in IC 10 are on the high end of dense gas SFE distribution with efficiencies similar to those seen in LIRGs and ULIRGs as well as in the outer disk of M51. } \label{fig:dense_gas_diag}
\end{figure*}

Having established in \S\S~\ref{sec:hcn_hcop} and \ref{sec:LIR-LHCN} that we can use \hcn\ and \hcop\ as dense gas tracers even in low metallicity sources like IC 10, Figure~\ref{fig:dense_gas_diag} shows the relationship between bulk molecular gas star formation efficiency (as traced by \LIRCO) and the dense gas fraction (as traced by either \LHCNCO\ or \LHCOPCO) along with points from our comparison samples. We find that the measured bulk molecular gas star formation efficiencies in IC 10 span the full range of molecular gas star formation efficiencies observed in other galaxies, i.e., one and a half orders of magnitude. The scatter in these points is consistent with the scatter shown in other resolved measurements in nearby galaxies and higher than the scatter for integrated measurements of entire galaxies (as one might expect based on the large amount of spatial averaging in the integrated measurements). 

The dense gas fractions for regions within IC 10 are on average lower than those found in LIRGs and ULIRGs. They are largely consistent with the dense gas fractions seen in a pointed survey of regions within nearby spiral galaxies \citep{Usero2015VariationsGalaxies} and with resolved measurements of the disk of M51 \citep{Bigiel2016THEM51}. The \LHCNCO\ ratio and limits for IC 10 are also consistent with values measured over similar size scales (${\sim} 34$~pc) for giant molecular clouds within the Milky Way \citep[$\LHCNCO=0.014\pm0.02$;][]{Helfer1997DenseWay}. We note that the dense molecular gas fractions and molecular gas star formation efficiencies measured by \citet{Braine2017DenseGalaxies} for a sample of low metallicity galaxies (including IC 10) occupy a similar location on the plots to our IC 10 points.

Uncertainties in our star formation rate and the \coh\ conversion factor may shift the location of the IC 10 points on this plot.  Note that if our star formation rates are underestimated by a factor of up to 4, the molecular star formation efficiencies in IC 10 shift towards the upper end of the range. However, metallicity-driven variations in the \coh\ conversion factor will tend to move points down by a factor of a few, leading to lower dense gas fractions, although the exact magnitude of the effect depends both on the \coh\ conversion factor and on how well \hcn\ and \hcop\ trace the dense gas at low metallicity. 

Point F is the exception to the above trends. It has an extremely high \LHCOPCO\ ratio, but its \LHCNCO\ limit is consistent with the other points within IC 10. We suggest that the high \LHCOPCO\ ratio is due to evolutionary effects rather than an intrinsically high dense gas fraction. If the dense gas fraction was intrinsically high, then we would expect both  \LHCOPCO\ and \LHCNCO\ to be high, with \LHCNCO\  likely higher than \LHCOPCO\ since \hcn\ traces gas with a higher critical density. This is clearly not the trend seen in panels (a) and (b) of Figure~\ref{fig:dense_gas_diag}.   We note that point F is located at the intersection of several gas bubbles in the central region of IC 10 (see Figure \ref{fig:spectra_overview} as well as \citealt{Wilcots1998The10} and \citealt{Leroy2006Molecular10}). \hcop\ emission can be enhanced by emission from shocks \citep{Viti1999ChemicalObjects}  The elevated \hcop\ emission at point F suggests that \hcop\ emission can be locally enhanced beyond what might expect from the gas density and thus \hcop\ may not always be a more reliable tracer of dense gas than \hcn, as suggested by \citet{Braine2017DenseGalaxies} and \citet{Johnson2018The210}.

We compare the dense gas fraction (as traced by \LHCNCO\ or \LHCOPCO) with the dense gas star formation efficiency (as traced by \LIRHCN\ or \LIRHCOP) for IC 10 and our sample of comparison galaxies in panels (c) and (d) of Figure~\ref{fig:dense_gas_diag}. The dense gas star formation efficiency in IC 10 is relatively high and is consistent with the high end of dense gas star formation efficiency found in individual pointings within galaxies \citep{Usero2015VariationsGalaxies,Bigiel2016THEM51} and the values found in LIRGs and ULIRGs \citep{Juneau2009,Garcia-Burillo2012Star-formationGalaxies}. The normal dense gas fractions and high dense gas star formation efficiencies are consistent with the values in the outer disk of M51 (c.f. Figures 3 and 4 in \citealt{Bigiel2016THEM51}), which has metallicity of  $12+\log(O/H)\sim 8.4$ \citep{Bresolin2004AbundancesM51}. They are also consistent with normal dense gas fractions and high dense gas star formation efficiencies found for a sample of low metallicity galaxies that includes IC 10 by \citet{Braine2017DenseGalaxies}.

This result assumes our \LIR-based star formation rates trace the true star formation in these galaxies.  If the star formation rates are a factor of 4 higher, then the dense gas star formation efficiencies in IC 10 would be on the high end of the distribution seen in other galaxies. As discussed above, point F has an anomalously high \LHCOPCO\ ratio that we attribute to evolutionary effects.

Figure~\ref{fig:dense_gas_diag} shows that the dense gas fractions  in IC~10 are low compared to those found in LIRGs and ULIRGs and comparable to those found in nearby spirals, while the dense gas star formation efficiencies are comparable to those found in LIRGs and ULIRGs and at the high end of nearby spiral galaxies. These results suggest that the CO and HCN emission occupy the same relative volumes as at higher metallicity, but that the size of the entire emitting region has been reduced.  This effect leads to normal dense gas fractions because the volume of the CO  and \hcn\ emitting regions have been reduced by a similar amount, but higher dense gas star formation efficiencies because the same star-forming region is associated with less \hcn\ or \hcop\ emission.  This picture is consistent with the results of \citet{Madden1997CGalaxies}, which showed that the [{\sc C\,ii}] and molecular hydrogen emission are largely coincident while the CO emission originates from a much smaller volume of the molecular cloud due to lack of self-shielding.  The structure of the emission implied by the dense gas fractions and dense gas star formation efficiencies in IC 10 implies that comparing dense molecular gas seen at higher redshift with that observed in the local universe will require a metallicity-based correction to compensate for a decrease in the volume of the molecular cloud traced by \hcn\ and \hcop.


\section{Summary} \label{sec:conclusions}

In this paper, we present new Green Bank Telescope observations of the dense molecular gas tracers \hcn\ and \hcop\ in the nearby low metallicity starburst galaxy IC 10 with the goal of quantifying the relationship between these tracers and star formation. Understanding how this relationship behaves in low metallicity systems provides key insights into observations of dense molecular gas at higher redshifts, where galaxies have lower metallicities.

We detect \hcn\ emission in one out of the five positions observed and \hcop\ emission in three of the five positions with a tentative fourth detection.  In all cases, the \IHCNHCOP\ ratio or upper limit on such a ratio is consistent with being less than one, similar to the ratio found in other low metallicity systems as well as the ratio in starburst galaxies. It is lower than the typical ratio found in the normal spiral galaxy M51. The observed \IHCNHCOP\ ratios in IC 10 could be the result of several different effects including the density distribution of the gas, abundance effects driven by star formation or lack of nitrogen, and/or shocks. Unfortunately, we do not have enough information to distinguish between these possibilities.  Regardless of the exact cause, the fact that the \IHCNHCOP\ values within IC 10 are within the range of values found for more massive galaxies suggests that changes in \IHCNHCOP\ due to metallicity are relatively small compared to other effects that contribute to \IHCNHCOP\ variations in more metal-rich systems. 

We compare our \LHCN\ and \LHCOP\ values to the infrared luminosity (\LIR) derived from infrared-color corrected 70~\micron\ Herschel data. Although in general \LIR\ may underestimate the star formation rate in low metallicity galaxies due to their relatively low dust content, the pointings observed here are bright in the infrared and the high resolution of the GBT biases our observations to dusty environments where \LIR\ is good tracer of star formation. We take as our systematic uncertainty in star formation the ratio between the \ha- and \LIR-derived total star formation rates for IC 10 (roughly a factor of 4). We find that our detections and limits are consistent with the $\LIR{-}\LHCN$ and $\LIR{-}\LHCOP$ relationships seen in integrated measurements of whole galaxies as well as resolved measurements in our own Milky Way and in other nearby galaxies. This result, combined with the relatively normal \IHCNHCOP\ ratios in this sytem, suggests that,  to first order,  \hcn\ and \hcop\  can be used as a dense gas tracers even in this low (1/4$Z_\odot$) metallicity system.

To gain insight into the relationship between dense gas and star formation in IC 10, we compare the dense molecular gas fraction, the molecular gas star formation efficiency, and the dense molecular gas star formation efficiency found in IC 10 to those in our comparison samples. The bulk molecular gas star formation efficiency in IC 10 is consistent with the full range of values observed in other galaxies. We find that the dense gas fraction in IC 10 is comparable to that seen in nearby galaxies, but lower than that in LIRGs or ULIRGS. However, the dense molecular gas star formation efficiency is relatively high and consistent with the upper end of values found in individual pointings within galaxies and in LIRGs and ULIRGs. If we have underestimated the star formation rate by a factor of a few, then both the bulk molecular gas and dense molecular gas star formation efficiencies for IC 10 would be even higher. The exception to these trends is pointing F, which has an anomalously high \LHCOP\ value. We suggest that the \hcop\ emission at this location, which is at the intersection of several molecular gas bubbles, has been enhanced due to shocks. 

The normal dense gas fractions, but high dense gas star formation efficiencies found in IC 10 suggest that the molecular gas in this low metallicity system retains the same relative structure for CO and HCN as in more metal-rich galaxies, but that the entire emitting structure has been pushed further into the cloud, reducing its overall size.  This implies that \hcn\ and \hcop\ will, in general, trace the dense molecular gas in high redshift galaxies, but that the conversion to the observed amount of dense molecular gas will likely need to be corrected for the metallicity of the source.

\facility{GBT, Herschel, CARMA}

\acknowledgments The authors thank an anonymous referee whose comments improved this paper. The National Radio Astronomy Observatory is a facility of the National Science Foundation operated under cooperative agreement by Associated Universities, Inc. The work of AKL is partially supported by the National Science Foundation under Grants No. 1615105, 1615109, and 1653300. This paper is partially supported by NSF grant 1413231 (PI: K. Johnson). F.B. acknowledges funding from the European Union's Horizon 2020 research and innovation programme (grant agreement No 726384 -- EMPIRE). AU acknowledges support from Spanish MINECO grants ESP2015-68964 and AYA2016-79006.


\end{document}